**Comment on "Unconventional s-wave superconductivity in Fe(Se,Te)"**

I.I. Mazin[1] and D.J. Singh[2]
[1] Naval Research Laboratory, Code 6390, 4555 Overlook Avenue Southwest, Washington, DC 20375, USA.
[2] Materials Science and Technology Division, Oak Ridge National Laboratory, Oak Ridge, TN 37831-6114, USA

**Abstract**

**Hanaguri et al (Reports, 8th of January 2010, p. 181)[1] report STM measurements on an Fe-based superconductor Fe(Se,Te). After Fourier-transforming their data they found three sets of sharp peaks in the reciprocal space. They interpreted one as a Bragg peak (an umklapp scattering) and two as resulting from quasiparticle interference (QPI), even though the observed peaks seem too sharp for QPI in a nodeless superconductor. However, at least one of these two peaks is also a Bragg peak, corresponding to another reciprocal lattice vector, and therefore is not a QPI peak and the third peak has similar structure suggesting an origin other than QPI.**

Hanaguri et al.[1] presented STM measurements of Fe(Se,Te) samples. They found, after Fourier-transforming their data, three sets of sharp peaks. One peak, located at the q-vectors equivalent to $2\pi/a,2\pi/a$, where $a=\sqrt{2}\ d_{Fe\text{-}Fe}$ is the *crystallographic* lattice parameter, was identified as a reciprocal lattice vector and discarded in the further analysis. In fact, in the Fe superconductors there are two Fe atoms per unit cell, which leads to a doubling of the density of Bragg vectors in reciprocal space. Because of this, another peak, labeled as $q_3$, which was interpreted as a QPI peak is also at a reciprocal lattice vector: $2\pi/a,0$ (in the supplementary materials, the authors suggest that this peak is an overlap of a QPI peak and a Bragg peak "of unknown origin", yet the entire peak is too sharp for a QPI feature, and the origin of a Bragg reflection is clear and purely crystallographic). The third peak, $q_2$, appears at the vector $\pi/a,\pi/a$, which is not a reciprocal lattice vector, but corresponds to the wave vector of the usual collinear spin density wave that competes with superconductivity in these materials. This peak is also sharp in momentum space similar to the other two suggesting a similar origin, i.e. a lattice ordering, in this case most likely on the surface.

Quasiparticle interference peaks are related to the momentum dependence of the order parameter on the Fermi surface. Therefore they have a characteristic momentum width that is related to this variation,[2] or, if there is no appreciable intraband variation, as in the case of the simple sign changing *s*-wave order, to the size of the individual Fermi surfaces. In the Fe-based superconductors the expected full width in this scenario is at least 15% and more reasonably 20% of the zone dimension, $2\pi/a$. On the other hand, features arising from long range periodicity, due to lattice or magnetic ordering, can be very sharp.

A closer look at the Hanaguri et al. data reveals that they cannot have a QPI origin whether one assumes a nodal (a) or an isotropic (b) order parameter. Indeed:

(a) If there is any gap anisotropy (in fact, direct measurements indicate that such

anisotropy exists and may be strong[3]), by far the strongest effect would be near-node scattering similar to the "octet" model in cuprates,[2,4] and, similar to cuprates, it would result in bias-dependent (position-wise), incommensurate peaks, rather that the observed bias-independent, commensurate peaks.

(b) If the gap is nevertheless isotropic, as conjectured in Ref. 1, the peaks in the QPI image corresponding to inter-band scattering would be only as sharp as the Fermi surface size (which is also the scale for the structure in the real part of the susceptibility). Indeed, we have calculated the QPI function Z, as defined in Eq. S5 from Ref. 4. The results are shown in Fig. 1 for different values of $q_z$. As expected, a complex pattern develops that extends over almost the entire Brillouin zone. At some particular values of $q_z$ four spots that are somewhat brighter than the rest appear that are located *not* at the M point, but at about 20% distance from it (towards the X points).

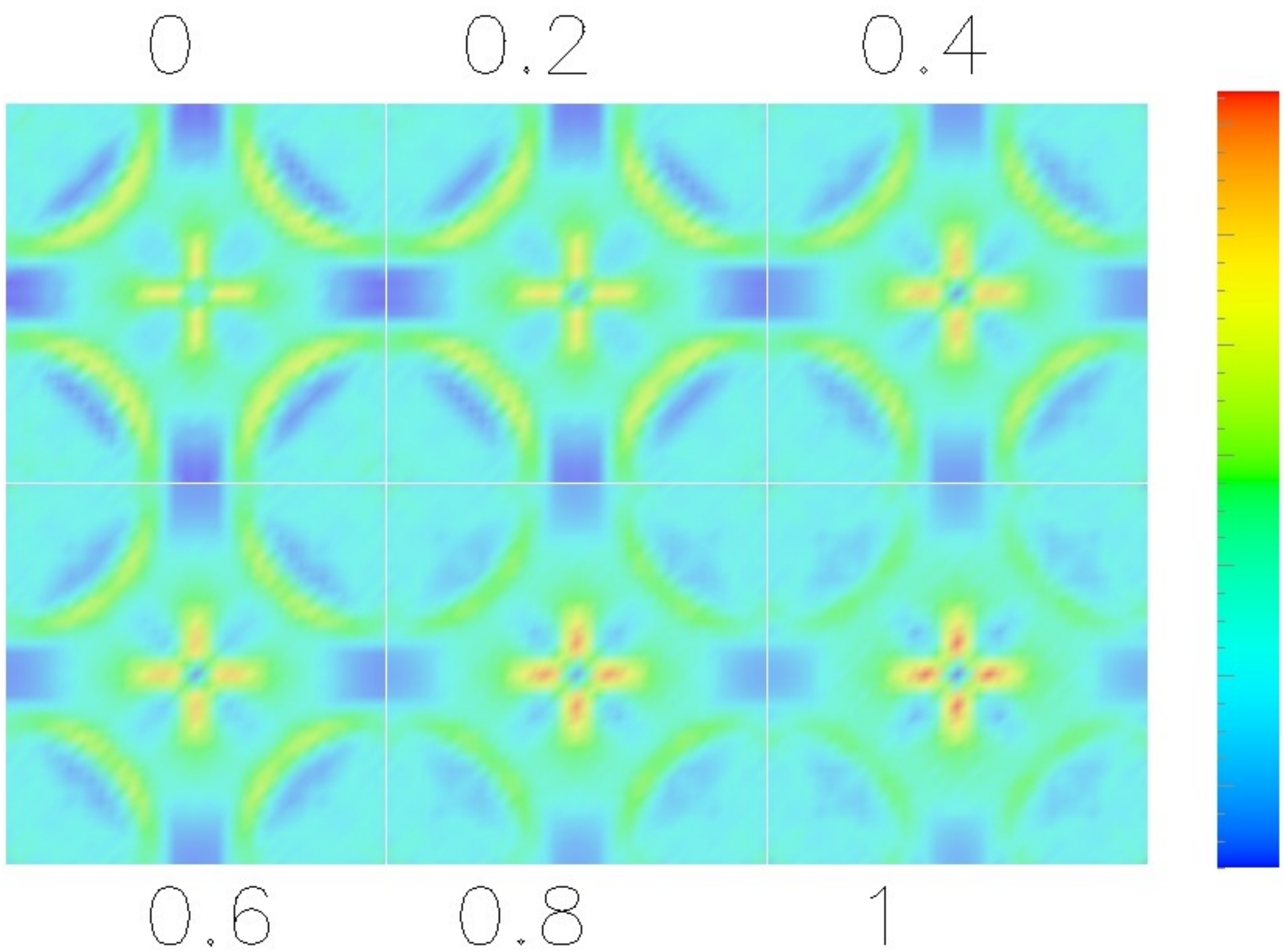

Figure 1. Calculated QPI spectra assuming a constant gap, for 6 values of $q_z$. The numbers label $q_z$ in units of $\pi/c$. The maximum value on the scale is about 2,5 larger than the minimal value.

This strongly argues against the interpretation in terms of QPI, and in favor of a surface reconstruction. Indeed, 2x1 and √2x√2 surface reconstructions have been found in other Fe-based superconductors, both experimentally and theoretically (see Ref. 5 for a discussion). A very interesting possibility is that the observed reconstruction is triggered by a coexisting magnetic order: it is known that the onset of the spin density wave (SDW) strongly affects the electron response function in other ferronictides, both in the experiment[6] and in the calculations[7]). It is worth noting that a magnetic order at $q_2=\pi/a,\pi/a$ was found to coexist in the bulk with superconductivity in $FeTe_{0.7}Se_{0.3}$ samples,[8] and also that a long range order with the same vector strongly emerges (and also coexists with superconductivity) in FeSe under small pressure.[9] Magnetic excitations with the $q_2$ wave vector are known to exist in Fe(Se,Te).

Interestingly, the latter scenario can explain why the peaks at $q_2$ and $q_3$ show different magnetic field dependencies. Indeed, since the measurements in question are done at biases comparable to the gap value, the pair-breaking effect of magnetic field affects the quasiparticle DOS and therefore the tunneling current, and correspondingly, enhances the intensity of the Bragg peaks. On the other hand, the magnetic field also affects the SDW formation, reducing its amplitude, and thus *reduces* the intensity of the Bragg peaks, if the surface reconstruction is triggered by the SDW. Experiments such as spin-polarized atomic resolution STM and spin polarized electron diffraction would be very helpful in addressing this interesting possibility. We want to emphasize, however, that while testing the “magnetic” scenario is a challenge, the fact remains that the observed features cannot be interpreted as the usual QPI peaks.

To conclude, we observe that (1) one of the two peaks interpreted as resulting from QPI by Hanaguri et al is actually a Bragg peak, (2) the other peak is too sharp and located at a high symmetry point, while QPI in this systems should invariably result in broad structures, very dissimilar to the observation, (3) this peak is consistent with a Bragg peak in a doubled unit cell, corresponding to the characteristic collinear SDW ordering in FeAs planes. We speculate that such SDW exists in the Fe(Se,Te) samples studied by Hanaguri et al, and triggers a 2x1 surface reconstruction with the same wave vector. This is consistent with the observed pattern and also could provide an explanation of the observed magnetic field dependence. Note that while one may, in principle, conjecture coexistence of a peak due to a surface ordering and a QPI peak at $q_2$, this is not an option for $q_3$, which corresponds to a reciprocal lattice vector of the underlying bulk crystal structure, and even at $q_2$ an overlapping QPI peak would have created a broad structure around the Bragg peaks rather than the observed small spots.

Work at ORNL was supported by the Department of Energy, Materials Science and Engineering Division. We are grateful for collegial discussions with Tetsuo Hanaguri and his sharing some unpublished data.